\documentclass[aip, jmp, amsmath, amssymb, cha]{revtex4-1}

\usepackage{amssymb}
\usepackage{amsmath}
\usepackage{multirow}

\usepackage{lineno}

\usepackage{graphicx}
\usepackage{caption}
\usepackage{subcaption}
\usepackage{filecontents}
\usepackage{natbib}

\begin{document}

\title{Quantification and Comparison of Degree Distributions in Complex Networks}

%
%
%
\author{Sadegh Aliakbary}
\email[]{aliakbary@ce.sharif.edu}
\affiliation{Department of Computer Engineering, Sharif University of Technology, Tehran, Iran}

\author{Jafar Habibi}
\email[]{jhabibi@sharif.edu.}
\affiliation{Department of Computer Engineering, Sharif University of Technology, Tehran, Iran}

\author{Ali Movaghar}
\email[]{movaghar@sharif.edu}
\affiliation{Department of Computer Engineering, Sharif University of Technology, Tehran, Iran}

\date{\today}

\begin{abstract}
The degree distribution is an important characteristic of complex networks. In many applications, quantification of degree distribution in the form of a fixed-length feature vector is a necessary step. On the other hand, we often need to compare the degree distribution of two given networks and extract the amount of similarity between the two distributions. In this paper, we propose a novel method for quantification of the degree distributions in complex networks. Based on this quantification method, a new distance function is also proposed for degree distributions, which captures the differences in the overall structure of the two given distributions. The proposed method is able to effectively compare networks even with different scales, and outperforms the state of the art methods considerably, with respect to the accuracy of the distance function. 

\end{abstract}



\maketitle 


\begin{quotation}
In many data-analysis applications, we should represent any network instance as a feature vector. Since the degree distribution is an important network feature, quantification (feature extraction) of the degree distribution plays an important role in these applications. On the other hand, we frequently need to compare two complex networks according to their degree distributions. The comparison is done in many applications, such as evaluation of network generation models and evaluation of sampling methods. In such applications, we frequently compare the degree distribution of two networks with different scales (different number of nodes/edges). The comparison of degree distributions is best achieved by the means of a scale-independent distance function. The current approaches for comparing degree distributions are Kolmogorov-Smirnov (KS) test, comparison based on fitted power-law exponent, and distribution percentiles method. Power-law exponent is too limited for comparing the distributions, particularly when they do not follow a power-law model. On the other hand, KS-test and percentile method are sensitive to the range of node degrees. In this paper, we propose a novel method for quantification and comparison of the degree distributions in complex networks. The proposed method, which is named Degree Distribution Quantification and Comparison (DDQC), considers the mean and the standard deviation of the node degrees to offer a scale-independent distance function.
\end{quotation}

\section{Introduction}
\label{Introduction}
There is a growing attention to the study of complex networks in recent years. Real-world networks such as social networks, biological networks and technological networks, display common topological features that discriminate them from random graphs \cite{ref1,ref2,ref3,ref4}. Among the features, small path lengths (small-world property), high clustering, community structure and heavy-tailed degree distribution are well studied in the literature \cite{ref2,ref3,ref4,ref5,ref6,ref7}. \\

The degree of a node specifies the number of its connections and the probability distribution of the degrees over the whole network forms the degree distribution. The degree distribution of complex networks often follows a heavy-tailed distribution \cite{ref8,ref9,ref10,ref11,ref12}, such as the power-law \cite{ref4,ref8,ref10} or the log-normal models \cite{ref13,ref14}. Although the degree distribution is an important network feature, the quantification and comparison of this feature is not a trivial task. We frequently need to compare degree distributions of complex networks. For example, in evaluation of sampling algorithms, we usually compare the given network instance with its sampled counterpart to ensure that the structure of the degree distribution is preserved \cite{ref11,ref15,ref16,ref17,ref18,ref19}. On the other hand, representing the network as a fixed-size feature vector is an important step in every data analysis process \cite{ref61,ref20,ref21}. To employ the degree distribution in such applications, a procedure is needed for extracting a feature vector from the degree distribution. The quantified feature vector is also useful in developing a distance function for comparing two degree distributions.\\

To the best of our knowledge, the current approaches for comparing degree distributions are eye-balling the distribution diagrams (usually, to satisfy a heavy-tailed distribution)  \cite{ref2,ref9,ref22}, Kolmogorov-Smirnov (KS) test  \cite{ref8,ref10,ref11,ref23,ref24,ref25}, comparison based on fitted power-law exponent \cite{ref22,ref26}, and comparison based on distribution percentiles \cite{ref20}. Eyeballing is obviously an inaccurate, error-prone and manual task. Comparison based on power-law exponent is based on the assumption that the degree distributions obeys a power-law, which is invalid for many complex networks \cite{ref13,ref14,ref38,ref39}. KS-test is based on a point-to-point comparison of the distributions, which is not a good approach for comparing networks with different ranges of node degrees. Percentile method is too sensitive to the outlier values of node degrees. As a result, the existing methods are actually inappropriate for comparing the degree distribution of networks, particularly when the target networks have different sizes and scales (e.g., when comparing the degree distribution of a large-scale and a small-size network).\\

We assume that considering the mean and standard deviation of the degree distribution in the quantification phase would make the comparison process more accurate and more robust, particularly with respect to scale variation. In this paper, we propose a new method called Degree Distribution Quantification and Comparison (DDQC) for this purpose. In our proposed ''quantification process'', a feature vector of some real numbers is extracted from the degree distribution, which can be used in data analysis applications, data-mining algorithms and comparison of degree distributions. In ''comparison task'', we aim a distance function that computes the distance (amount of dissimilarity) between two given network degree distributions. With such a distance function, we can figure out how similar the given networks are, according to their degree distributions. The distance function should return small values for similar networks and large values for dissimilar networks. As an evaluation of our proposed method, we compare it with existing baseline methods. The evaluation shows that the proposed method offers an effective quantified representation of the degree distributions and outperforms the baseline methods with respect to the accuracy of the distance function. Although our proposed approach is of general nature and equally applicable to other network types, in this paper we focus on simple undirected networks.\\

The rest of this paper is organized as follows: The section \ref{Motivation}  describes the motivation of this research, along with its applications. In section \ref{Related Works}, we briefly overview the related works. In section \ref{Proposed Method}, we propose a new method for degree distribution quantification and comparison. In section \ref{Evaluation}, we evaluate the proposed method and we compare it with baseline methods. Finally, we conclude the paper in section \ref{Conclusion}.\\

\subsection{Motivation}
\label{Motivation}
Degree distribution is an important and informative characteristic of a network \cite{ref1,ref2,ref3,ref4,ref7,ref8,ref9,ref10,ref11,ref13,ref16,ref19,ref22, ref27,ref28,ref29,ref30}. Although it does not capture all aspects of the topology of a network \cite{ref18}, it reflects the overall pattern of connections \cite{ref27} and is an important determinant of network properties \cite{ref28}. Degree distribution is also a sign of link formation process in the network. For example, preferential attachment process results in a power-law degree distribution \cite{ref7}, and log-normal degree distributions imply that there are probabilistically more low degree social nodes in the network than those in power-law distributed networks \cite{ref6}. Although it is hard to detect the exact link formation process only based on the degree distribution, a similarity metric for degree distribution is certainly helpful in this task, along with other network features such as average path length \cite{ref3}, clustering coefficient \cite{ref5}, and community structure \cite{ref6}. This similarity metric plays an important role in evaluation of network generation models \cite{ref1,ref2,ref5,ref7,ref9,ref29}, evaluation of sampling methods \cite{ref11,ref15,ref16,ref17,ref18}, generative model selection \cite{ref61,ref20,ref21}, and many other applications.\\

Network generation is an area of research, in which the degree distribution of the network is considered. Generative models, such as Watts-Strogatz model \cite{ref5}, Barab\'{a}si-Albert model \cite{ref7} and Kronecker graphs model \cite{ref29}, synthesize realistic networks that mimic the properties of real-world networks, including small-world property, high clustering and heavy tailed degree distribution. Many generative models are able to fit for a target network instance. For example, KronFit tries to find parameters that well mimic the properties of the given target network, for Kronecker graphs model \cite{ref29}. For evaluating a generative model, the proposed distance function plays an important role in comparing the generated networks with the target network according to their degree distributions. In this application, the generated network may have a scale different from the original given network. So, the distance metric should be able to compare networks of different sizes and should be able to capture the similarities between networks of different scales (e.g., a large network and a small network instance). The need to a quantitative similarity measure becomes more important when we consider various generative models that produce similar distributions. For instance, Barab\'{a}si-Albert \cite{ref7}, Forest Fire \cite{ref9}, Kronecker graphs \cite{ref29}, and random power-law \cite{ref30} models all generate networks with heavy-tailed degree distributions.\\

With the current growing number of generative models in the literature, an important pre-stage of network generation is to identify the generative model that best fits to the target network. This problem is called generative model selection \cite{ref61,ref19,ref20,ref21,ref31}. Proposed solutions for this problem study the features of the target network and find the best generative model that reproduces these features. Undoubtedly, an appropriate solution should encounter the degree distribution as one of the network features. In this context, our demanded quantification and similarity metric for degree distributions is applicable.\\

There is another track in network research aiming for “sampling” from large networks \cite{ref11,ref15,ref16,ref17,ref18,ref19}. Sampling methods try to extract a random subset of a large network while keeping the degree distribution, diameter and other connectivity patterns of the original network in the sampled graph \cite{ref19}. The desired distance metric of this paper can be used for evaluating a sampling method by specifying how similar the degree distribution of a sampled network is to the original network. Again, in this application we should compare networks of different scales: a large network is compared with a sampled smaller network.\\

Comparison of the networks based on their structural features, including the degree distribution, is essential in other applications such as classification or clustering of network instances \cite{ref3,ref19,ref20,ref21,ref61}, anomaly detection \cite{ref32,ref33}, and study of epidemic dynamics \cite{ref34,ref35,ref36}. It is also worth noting that feature extraction (quantification) from the degree distributions has important applications, even if no network comparison is required. Particularly, data-analysis algorithms require the network features, including the degree distribution, to be represented as some real numbers in the form of a fixed-length feature vector \cite{ref61,ref19,ref20,ref21}.\\

According to the mentioned applications for the demanded distance function, we regard two degree distributions similar if their networks follow similar structure, similar connection patterns and similar link formation processes, even if the networks are of completely different scales and sizes. As an example, consider the similarity of a temporal network (a network with a known timestamp) to its consequent snapshot over time. If the link formation process is relatively stable over the time, we expect a temporal network to be similar to other near-in-the-time networks. For example, the Facebook social network at 2006 is similar to that network at 2005 or 2007, and this network is meaningfully less similar to a completely different network, such as the citation network of DBLP \cite{ref39}. In this paper, we try to capture such similarities based on the degree distribution of the networks.\\

\section{Related Works}\label{Related Works}
The degree distribution of many real-world networks are heavy tailed \cite{ref8,ref9,ref10,ref11,ref12}, with the power-law distribution as the most suggested model \cite{ref4,ref8,ref10}. The power-law distribution is also observed in eigenvalues and eigenvectors \cite{ref29}, densification \cite{ref9} and other networks features. In power-law degree distribution the number of nodes with degree d is proportional to $d^{-\gamma}$ ($ N_{d} \varpropto d^{-\gamma} $) where $\gamma$ is a positive number called ''the power-law exponent''. The value of $\gamma$ is typically in the range $2 < \gamma < 3$ \cite{ref2,ref37,ref38}. The exponents of the fitted power-law can be used to characterize graphs \cite{ref26}. A common approach for quantifying the degree distribution is to fit a power-law distribution on the network distribution and to find its power-law exponent ($\gamma$). This quantity is comparable for different networks and it is possible to compare networks according to their fitted power-law exponents. The problem with this approach is that power-law exponent is a single number and it is too limited to represent a whole degree distribution. This approach also follows the assumption that the degree distribution is power-law, which is not always valid, because many networks follow other degree distribution models  such as  log-normal distribution  \cite{ref13,ref14,ref38,ref39}. In addition, the power-law exponent does not reflect the deviation of the degree distribution from the fitted power-law distribution. As a result, two completely different distributions may have similar quantified feature (fitted power-law exponent).\\

Degree distribution is a kind of probability distribution and there are a variety of measures for calculating the distance between two probability distributions. In this context, perhaps the most common approach is the Kolmogorov-Smirnov (KS) test, which is defined as the maximum distance between the cumulative distribution functions (CDF) of the two probability distributions \cite{ref8}. KS-test is used for comparing two degree distributions (two-sample KS test) \cite{ref11,ref23} and also for comparing a degree distribution with a baseline (usually the power-law) distribution \cite{ref8,ref13,ref40,ref41}. The KS distance of two distributions is calculated according to Equation \ref{eq:KS}, in which $S_{1}(d)$ and  $S_{2}(d)$ are the CDFs of the two degree distributions, and $d$ indicates the node degree. KS-test is largely utilized in the literature for comparing degree distribution of networks \cite{ref8,ref10,ref11,ref23,ref24,ref25}. KS-test is a method for comparing the degree distributions and calculating their distance, and it does not provide quantification or feature extraction mechanism. We need to maintain the CDF of the degree distributions so that we can compare them according to KS-test. KS-test is also sensitive to the scale and size of the networks, since it performs a point-to-point comparison of CDFs. So, for two networks of completely different scales (e.g., a very large network and a relatively small network, with different ranges of node degrees), the KS-test will probably return a large value as their distance, even if the overall views of the degree distributions are similar.\\

\begin{equation} \label{eq:KS}
distance_{KS}(S_{1},S_{2})=\max_{d}\vert S_{1}(d)-S_{2}(d) \vert
\end{equation}

Janssen et. al., \cite{ref20} propose another approach for quantification of degree distributions. In this method, the degree distribution is divided into eight equal-sized regions and the sum of degree probabilities in each region is extracted as distribution percentiles. This method is sensitive to the range of node degrees and also to outlier values of degrees. We recall this technique as ''Percentiles'' and we include it in baseline methods, along with ''KS-test'' and ''Power-law'' (the power-law exponent) to evaluate our proposed distance metric, called ''DDQC''. \\

\section{Proposed Method}
\label{Proposed Method}

We propose a new method for quantifying and comparing the degree distribution of networks. In this method, a vector of at least four real numbers is extracted from the degree distribution. A distance function is also suggested for comparing the quantified vectors. As discussed above, an appropriate distance metric for degree distributions should be able to effectively compare networks, even if they have different range of node degrees. To eliminate impact of the network size from the quantification of its degree distribution, we considered the mean and standard deviation of the degree distribution in the quantification procedure. The following two subsections show our proposed method for quantification and comparison of degree distributions.\\

\subsection{Quantification of Degree Distribution}
\label{Quantification of Degree Distribution}

\begin{figure*}
\begin{center}
\includegraphics{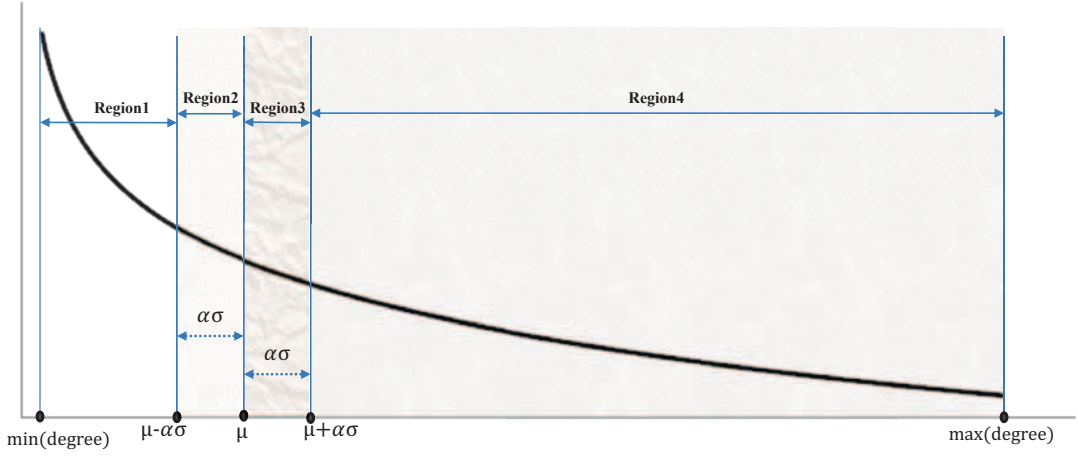}
\caption{ \label{fig:regions} The four regions of the degree distribution. According to the mean and standard deviation of node degrees, the distribution is divided into four regions and the interval probabilities are computed for the regions and the more fine-grained intervals.}
\end{center}
\end{figure*}

The degree distribution of a network is described in Equation \ref{eq:PG} as a probability distribution function. The aim of ''quantification'' task is to extract a fixed-length vector of real numbers as the representative of the degree distribution. Then we can use this vector in network analysis and also network comparison. In the first step of quantification, as Figure \ref{fig:regions} shows, we define four regions in the degree distribution of a given network. These regions are separated based on five ''region points'': $min(degree)$, $\mu - \alpha \sigma$, $\mu$, $\mu + \alpha \sigma$ and $max(degree)$. $min(degree)$ is the minimum of all the existing degrees in the degree distribution, $\mu$ is the mean of degrees according to their probabilities (Equation \ref{eq:meanG}) , $\sigma$ is the standard deviation of the degrees (Equation \ref{eq:stdxG}), $\alpha$ is a configurable parameter (it specifies the width of the regions), and $max(degree)$ is the maximum existing node degree in the network. The smallest possible feature vector in our proposed method is a vector of four numbers, each of which showing the sum of the probability of degrees in one of the four specified regions. For finer comparison of distributions, we can further divide each region into $L$ equal-size intervals. In our experiments, $L$ is set to $2^\beta$, where $\beta$ is a positive integer value ($\beta \geqslant 0 $) and the second configurable parameter of our method. Larger values of $\beta$ results in a more fine-tuned quantification and also more elements in the feature vector.  While even small values for $\beta$ (e.g.,  $\beta =1$) brings a more accurate distance metric compared with the baseline methods, larger values of $\beta$ improves the accuracy of the distance function. So, tuning $\beta$ parameter is a tradeoff between the accuracy of the algorithm and the size of the quantified feature vector.\\

According to Figure \ref{fig:regions}, the four regions in the degree distribution ($region_G(r) , r=1..4$) are illustrated in Equation \ref{eq:regions}. Equation \ref{eq:regionlength} shows the length of each region ($\vert region_G (r)\vert$), which is equal to the absolute difference between the region endpoints. Each region is then divided into $L=2^\beta$ equal-length intervals. Equation \ref{eq:intervalpoints} shows the interval borders ($intervalpoint_G(b,L)$) and Equation \ref{eq:intervals} shows the defined ranges for intervals ($interval_G(i,L)$), in which $b$ is the interval point counter, $L=2^\beta$ is the split factor of each region, and $i$ is the interval identifier. The ''interval degree probability ($IDP_G$)'' is defined in Equation \ref{eq:idp} as the sum of degree probabilities in a specified interval. Actually, the ''less-than'' comparator in Equation \ref{eq:idp} becomes ''less-than or equal'' only for the last (rightmost) interval. Equation \ref{eq:quantification} shows the final quantified feature vector, which contains $4L=4\times2^\beta$ elements, each of which is the $IDP$ for one of the defined intervals. In this equation, $quantification_{\beta}(G)$ is the quantified representation of the degree distribution as a fixed-length feature vector with $4\times2^\beta$ elements. It is worth noting that if the upper bound of a region is turned out to be less than its lower bound, all the degree probabilities ($IDP$s) of that region will be set equal to zero. For example, in the first region, $region_{G}(1)=\left[ min(degree(v)), \mu-\alpha \sigma \right] $, if $min(degree(v)) > \mu-\alpha \sigma$, then all probabilities in the first region will be considered equal to zero. This situation is possible only for the first and the last (fourth) regions, since the size of the second and the third regions are always positive values.\\

\begin{equation} \label{eq:PG}
P_G (d)= P(degree(v)=d );v\in V(G)
\end{equation}

\begin{equation} \label{eq:meanG}
\mu _G =\sum\limits_{d=min_G (degree(v))}^{max_G (degree(v))} d \times P_G (d)
\end{equation}

\begin{equation} \label{eq:stdxG}
\sigma _G =\sqrt{\sum\limits_{d=min_G (degree(v))}^{max_G (degree(v))}  P_G (d)\times(d-\mu _G)^2}
\end{equation}

\begin{equation}
\label{eq:regions}
region_G (r) = 
\begin{cases} 
\left[ min_G(degree(v)), \mu_G - \alpha  \sigma_G \right]  &\mbox{if } r=1 \\
\left[ \mu_G - \alpha  \sigma_G , \mu_G \right]  & \mbox{if } r=2 \\
\left[ \mu_G , \mu_G + \alpha  \sigma_G \right] & \mbox{if } r=3 \\
\left[ \mu_G + \alpha  \sigma_G  , max_G(degree(v))\right] & \mbox{if } r=4.

\end{cases}
\end{equation}

\begin{equation} \label{eq:regionlength}
\vert region_G (r)\vert = length(region_G (r))=⁡max(max(region_G (r)) - min(region_G (r)),0)
\end{equation}

\begin{equation}
\label{eq:intervalpoints}
intervalpoint_G (b,L) = 
\begin{cases} 
min_G⁡(degree(v))+ \frac{(b-1)\times \vert region_G (1)\vert}{L}   &\mbox{if } 1 \leq b \leq L \\
\mu _G - \alpha\sigma _G+ \frac{(b-L-1)\times \vert region_G (2)\vert}{L}  & \mbox{if } L+1 \leq b \leq 2L \\
\mu _G + \frac{(b-2L-1)\times \vert region_G (3)\vert}{L}  & \mbox{if } 2L+1 \leq b \leq 3L \\
\mu _G + \alpha\sigma _G + \frac{(b-3L-1)\times \vert region_G (4)\vert}{L}  & \mbox{if } 3L+1 \leq b \leq 4L+1.

\end{cases}
\end{equation}

\begin{equation} \label{eq:intervals}
interval_G (i,L)=\left[ intervalpoint_G (i,L),intervalpoint_G (i+1,L)\right] ; i=1..4L
\end{equation}

\begin{equation} \label{eq:idp}
IDP_G (interval)=P(min(interval)\leq degree(v) < max(interval));   v\in V(G)
\end{equation}

\begin{equation} \label{eq:quantification}
quantification_\beta (G)=\left\langle IDP_G (interval_G (i, 2^\beta)) \right\rangle _{i=1..4\times 2^\beta}
\end{equation}

\subsection{Comparison of Degree Distributions}
\label{Comparison of Degree Distributions}
After the quantification task, we can compare the degree distribution of two networks $G_1$ and $G_2$ according to their quantified feature vectors. We assume that the two degree distributions are quantified with the same configuration parameters of $\alpha$ and $\beta$. Since the $\beta$ parameter is considered equal for both the networks, the size of the quantified vectors $quantification_\beta (G)$ will be equal for the two networks. For small values of $\beta$, $quantification_\beta (G)$ will show a coarse-grained representation of the degree distribution with few real numbers. For larger values of $\beta$, more fine-grained intervals of the degree distribution are available. According to Equation \ref{eq:idplevels}, we can simply compute the elements of $quantification_\beta (G)$ based on the elements in $quantification_{\beta +1} (G)$. In other words, it is possible to calculate $IDP_G$ for smaller values of $\beta$ (coarse-grained quantification) using $IDP_G$ with larger values of $\beta$ (fine-grained quantification).\\

Finally, we propose the Equation \ref{eq:distance} for comparing two degree distributions. This equation compares two networks based on their corresponding $IDP_G$ values for different granularities, from larger intervals (with $s=0$) to smaller intervals (with $s=\beta$). A coefficient ($\gamma^s$), which is the third configurable parameter of our framework, is also included to influence the impact of different granularities. Intuitively, $d(G_1 , G_2 )$ compares the corresponding interval degree probabilities of the two networks, sums their differences, and also includes a discount factor of $\gamma$ for the more fine-granularity intervals to raise the impact of course-grained intervals. Equation \ref{eq:distance} is a distance function for degree distribution of networks, and it is the result of a comprehensive study of different real, artificial and temporal networks.\\

\begin{equation} \label{eq:idplevels}
IDP_G (interval_G (i,L)=IDP_G (interval_G (2i-1,2L)+IDP_G (interval_G (2i,2L)
\end{equation}

\begin{equation} \label{eq:distance}
d(G_1,G_2)=distance(G_1,G_2)=\\
	\sum\limits_{s=0}^{\beta} \gamma ^s \sum\limits_{i=1}^{4\times 2^s} 
	\vert IDP_{G_1} (interval_{G_1} (i,2^s) - IDP_{G_2} (interval_{G_2} (i,2^s) \vert
\end{equation}

\section{Evaluation}
\label{Evaluation}
In this section, we comprehensively evaluate our proposed method. In section \ref{Datasets} we discuss the different network datasets which are used in our evaluations. In section \ref{Evaluation Criteria} we study the evaluation criteria, and finally in section \ref{Evaluation Results}, we compare our proposed method with baseline methods.

\subsection{Datasets}
\label{Datasets}
In our problem setting, we aim a distance function that given the degree distribution of two networks, calculates how similar they are. But what does this ''similarity'' mean for degree distributions? What benchmark is available for evaluating such a distance function? According to the motivation and the described applications for the demanded distance function, we regard two degree distributions similar if their networks follow similar link formation processes, even if the networks are of completely different scales and sizes. For evaluating different distance metrics, an approved dataset of networks with known distances of its instances is sufficient. Although there is no such an accepted benchmark of networks with known ''distance values'', there exist some similarity witnesses among the networks. For evaluating different distance metrics, we have prepared three network datasets with admissible similarity witnesses among the networks of these datasets:
\begin{itemize}
\setlength{\itemindent}{-1em}
\item 
\textbf{\textit{Artificial Networks}}. 
We have generated 8,000 artificial networks using eight generative models (1,000 network instances for each generative model). The selected generative models are Barab\'{a}si-Albert model \cite{ref7}, copying model \cite{ref42,ref43}, Erd\H{o}s-R\'{e}nyi \cite{ref44}, Forest Fire \cite{ref9}, Kronecker model \cite{ref29}, random power-law \cite{ref30}, Small-world (Watts–Strogatz) model \cite{ref5}, and regular graph model. For each generative model, 1,000 network instances are generated using completely different parameters. The number of nodes in generated networks ranges from 1,000 to 5,000 nodes with the average of 2,936.34 nodes in each network instance. The average number of edges is 13,714.75. In this dataset, the generative models (generation methods) are the witnesses of the similarity: the networks generated from the same model follow identical link formation rules, and their degree distributions are considered similar. The networks of this data-set are described in the \ref{datasetsappendix}, along with an overview of the selected generative models.
\item
\textbf{\textit{Real-world Networks}}. 
We have collected a dataset of 33 real-world networks of different types, most of them are publicly available in the web. The networks are selected from six different network classes: Friendship networks, communication networks, collaboration networks, citation networks, peer to peer networks and graph of linked web pages. The category of networks is a sign of similarity: networks of the same type usually follow similar link formation procedures and produce similar degree distributions. So, when comparing two network instances, we expect the distance metric to return small distances (in average) for networks of the same type and relatively larger distances for networks with different types. The ''real-world networks'' data-set is described in the \ref{datasetsappendix}, along with the basic specifications and the source of its networks.
\item
\textbf{\textit{Temporal Networks}}. 
We have access to temporal versions of two instances of the real-world networks dataset: Cit\_CiteSeerX, which is a citation network extracted from CiteSeerX digital library \cite{ref45} and Collab\_CiteSeerX which is a collaboration network (co-authorship) obtained from the same service. For each of the two temporal networks, we extracted nine snapshots of the network from 1994 to 2010 biannually (1994, 1996, …, 2010). For example, ''Cit\_CiteSeerX\_2010'' is a citation network of the papers published before 2010 which is extracted from CiteSeerX. It is reasonable to assume that the link formation in ''Cit\_CiteSeerX'' citation network follows a stable (not rapidly changing) process. For instance, the ''Cit\_CiteSeerX\_2008'' snapshot, is regarded similar to the ''Cit\_CiteSeerX\_2006'' and ''Cit\_CiteSeerX\_2010'' snapshots, and meaningfully less similar to other network instances in the real-worlds dataset. A similar argument is also valid for Collab\_CiteSeerX networks.

\end{itemize}

\subsection{Evaluation Criteria}
\label{Evaluation Criteria}
In the section \ref{Datasets}, we described our three network datasets and we introduced different signs and witnesses of similarities among networks of these datasets. We can consider these witnesses in the evaluation of the proposed method. We evaluate our proposed distance function and we compare it with baseline methods based on their consistency to mentioned witnesses of the similarity. For this purpose, we consider the following measurable criteria:

\begin{itemize}
\setlength{\itemindent}{-1em}
\item 
\textbf{\textit{kNN-Accuracy}}. 
The k-Nearest-Neighbor rule (kNN) \cite{ref46} is a common method for classification. It categorizes an unlabeled example by the majority label of its k-nearest neighbors in the training set. The performance of kNN is essentially dependent on the way that similarities are computed between different examples. So, better distance metrics result in better classification accuracy of kNN. To evaluate the accuracy of different distance functions, we employ them in kNN classification and we test the accuracy of this classifier. This evaluation is performed for both datasets of real-world and artificial networks. 
\item 
\textbf{\textit{Inter-class and intra-class distances}}. 
An appropriate distance metric should return smaller distances for networks that are chosen from the same class. In other words, when there exist predefined classes of networks, the distance metric is expected to report a small distance between two classmate networks and large distance between two networks of different classes. To evaluate a distance metric based on this requirement, we calculate the distance between any pair of networks of a dataset and we check the distance among classmate instances to be relatively smaller. This experiment is separately studied for both real-world networks and artificial networks datasets. In order to compare different distance metrics, we normalize distances of each distance metric according to its mean and standard deviation. As Equation \ref{eq:zscore} shows, the \textbf{z-score} (also called standard score, z value and normal score) is used for normalizing distance values. In this formula, $\mu _{S,d}$ shows the average pairwise-distances for networks in $S$ dataset, according to the $d$ distance metric (Equation \ref{eq:mu}), and $\sigma _{S,d}$ indicates the standard deviation of the pairwise-distances (Equation \ref{eq:sigma}). $nd_{S,d} (G_1 , G_2)$ shows the normalized distance between $G_1$  and $G_2$ networks based on the population of $S$ dataset of networks and $d$ distance metric. Normalized distance ($nd$) is an appropriate base for evaluating the accuracy of distance metrics, since it is a dimensionless quantity and shows the number of standard deviations that a distance is above the average distance. Z-score is widely used in the literature for similar purposes \cite{ref3,ref37,ref59,ref60}. The normalization emphasizes the relative magnitude of the distances rather than their absolute magnitude, which is important for the comparison of computed distances in different distance functions. The average of normalized distances ($nd$) in a dataset is equal to zero, similar instances result a small (negative) normalized distance and dissimilar instances show large (positive) normalized distances. Equation \ref{eq:intra} defines the average of normalized intra-class distances ($INTRA_d$) and Equation \ref{eq:inter} defines the average of normalized inter-class distances ($INTER_d$). Since $INTRA_d$ indicates the distance among networks of the same class, an accurate distance function results in a small negative value of $INTRA_d$. On the other hand, $INTER_d$ shows the distance among networks with different classes, and an appropriate distance function should indicate a large positive $INTER_d$ value.\\

\begin{equation} \label{eq:mu}
\mu _{S,d}=\dfrac{1}{\vert S \vert \times (\vert S \vert -1)}
\sum\limits_{G_1 , G_2 \in S , G_1 \neq G_2} d(G_1 , G_2)
\end{equation}

\begin{equation} \label{eq:sigma}
\sigma _{S,d}=\sqrt{
\dfrac{1}{\vert S \vert \times (\vert S \vert -1)}
\sum\limits_{G_1 , G_2 \in S , G_1 \neq G_2} (d(G_1 , G_2) - \mu _{S,d})^2
}
\end{equation}

\begin{equation} \label{eq:zscore}
nd_{S,d}(G_1 , G_2)=\dfrac{d(G_1 , G_2) - \mu _{S,d}}{\sigma _{S,d}}
\end{equation}

\begin{equation} \label{eq:intra}
INTRA_{S,d}=average(nd_{S,d} (G_1,G_2)) ; G_1,G_2\in S,class(G_1)=class(G_2)
\end{equation}

\begin{equation} \label{eq:inter}
INTER_{S,d}=average(nd_{S,d} (G_1,G_2)); G_1,G_2\in S,class(G_1)\neq class(G_2)
\end{equation}

\item 
\textbf{\textit{Distances of Temporal Networks}}. 
In a temporal dataset of networks, a network snapshot is expected to be similar to its neighbor snapshots. To test this requirement, we separately considered two datasets of temporal networks (Cit\_CiteSeerX\_2010 and Collab\_CiteSeerX\_2010) along with the instances of real-world networks dataset. We calculate pairwise distances for all the instances in the two aggregated datasets and then we normalize the distances according to Equation \ref{eq:zscore}. For a temporal network instance, the average of its normalized distances to the neighbor snapshots is expected to be a small value. This is because the most similar networks to a network snapshot are the neighbor snapshots of the same network.
\end{itemize}

Using the specified criteria, we compare our proposed method with three existing baseline methods which are described in the related works: ''Power-law'', ''KS-test'' and ''Percentiles''. It is worth noting that ''KS-test'' actually does not include a quantification mechanism and needs the whole degree distributions to operate. This is a drawback of KS-test, since other baseline methods and our proposed distance metric need a small quantification of the degree distributions (e.g., a feature vector of eight real numbers) for comparing two networks. On the other hand, ''Percentiles'' method is only a quantification algorithm and it offers no distance function. In our evaluations, we used Manhattan distance (also called city block and taxicab geometry) \cite{ref47} as the distance function for ''Percentiles'', because it resulted in better accuracy, compared with Euclidean distance and some other metrics.

\subsection{Evaluation Results}
\label{Evaluation Results}
In this subsection, we comprehensively evaluate the proposed method (DDQC) and compare it with the baseline methods. As described in section \ref{Proposed Method}, the proposed method is configurable by three parameters: $\alpha$, $\beta$ and $\gamma$. We start the evaluations by setting $\alpha =1$, $\beta =1$ and $\gamma =0.8$.\\

\begin{figure}
\begin{subfigure}[b]{0.45\textwidth}
	\includegraphics{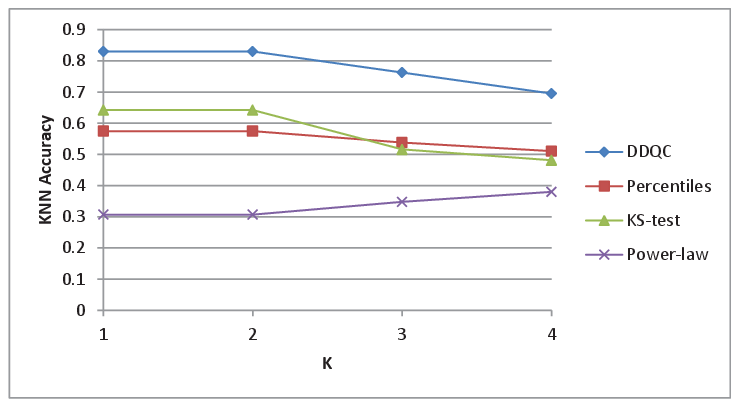}
	\caption{ \label{fig:knnart} kNN Accuracy in artificial networks dataset}
	\end{subfigure}
\begin{subfigure}[b]{0.45\textwidth}
	\includegraphics{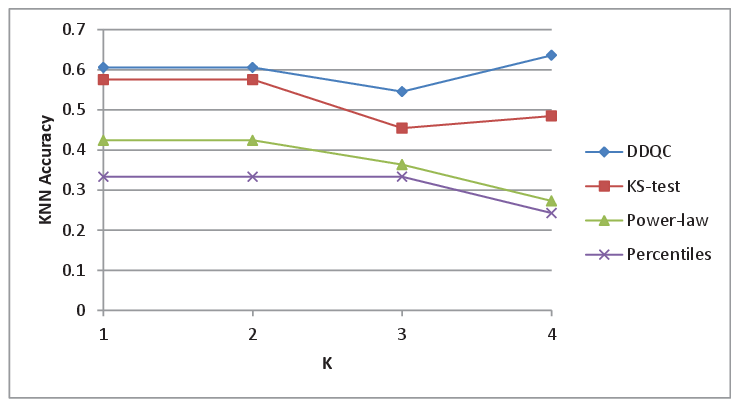}
	\caption{ \label{fig:knnreal} kNN Accuracy in real-world networks dataset}
	\end{subfigure}

	\caption{ \label{fig:knn} kNN Accuracy for different methods. When DDQC (here, with $\alpha =1, \beta =1, \gamma=0.8$) is employed as the distance function in the kNN algorithm, the accuracy of the classifier is the best, in both artificial (\subref{fig:knnart}) and real-world (\subref{fig:knnreal}) networks datasets, and for various values of $K$. }		
\end{figure}

For evaluating kNN accuracy on artificial networks dataset (Figure \ref{fig:knnart}), we iteratively created a small subset of this dataset and performed kNN on all instances of the formed subset. In each iteration of this experiment, we randomly selected 50 network instances from the dataset and computed the kNN accuracy for the set of these instances. Figure \ref{fig:knnart} shows this evaluation and reports the average of kNN accuracy on 100 independent iterations, for several values of $K$.  Figure \ref{fig:knnreal} shows the evaluation of different methods based on their kNN accuracy for real-world networks dataset. In this experiment, 33 real network instances, with known class labels, are classified using kNN algorithm and the average accuracy of the classifier is measured. According to Figures \ref{fig:knnart} and \ref{fig:knnreal}, DDQC outperforms all the baseline methods considerably with respect to kNN-accuracy, in both datasets of real networks and artificial networks. The evaluations are performed for different values of $K$ to ensure that the superiority of DDQC is not dependent on a particular $K$ value.\\ 

\begin{figure}
\begin{subfigure}[b]{0.45\textwidth}
	\includegraphics{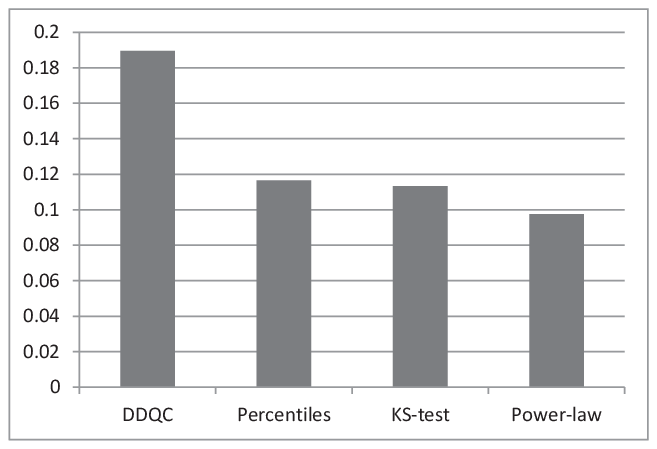}
	\caption{\label{fig:artinter} $INTER_d$ for artificial networks}
\end{subfigure}
\begin{subfigure}[b]{0.45\textwidth}
	\includegraphics{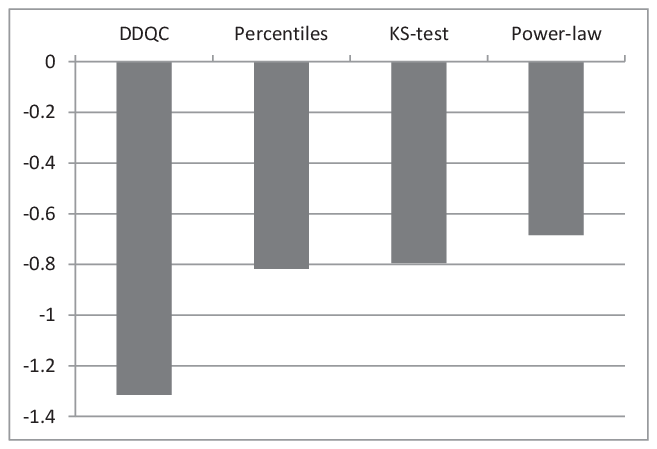}
	\caption{\label{fig:artintra} $INTRA_d$ for artificial networks}
\end{subfigure}
\begin{subfigure}[b]{0.45\textwidth}
	\includegraphics{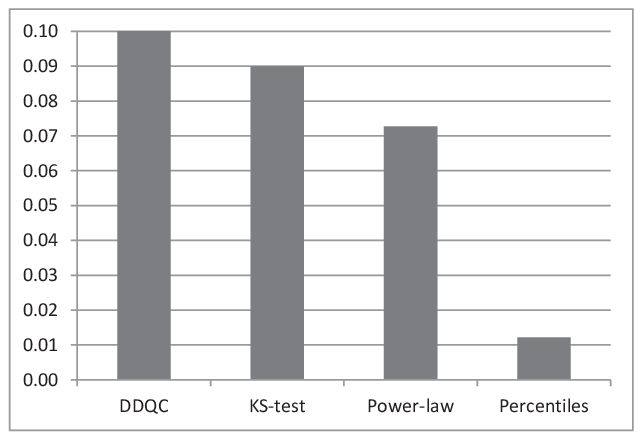}
	\caption{\label{fig:realinter} $INTER_d$ for real-world networks}
\end{subfigure}
\begin{subfigure}[b]{0.45\textwidth}
	\includegraphics{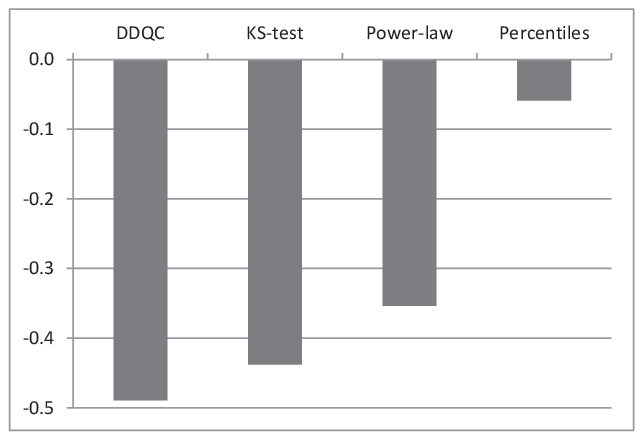}
	\caption{\label{fig:realintra} $INTRA_d$ for real-world networks}
\end{subfigure}

\caption{\label{fig:interintra} $INTRA_d$ and $INTER_d$ for artificial and real-world datasets. DDQC shows the largest normalized inter-class distances and the smallest normalized intra-class distances, in both datasets.}
\end{figure}

In the next experiment, we evaluate different methods based on $INTRA_d$ (Equation \ref{eq:intra}) and $INTER_d$ (Equation \ref{eq:inter}). This experiment is performed separately for real-world networks and artificial networks datasets and the results are illustrated in Figure \ref{fig:interintra}. As the figures indicate, DDQC outperforms all the baseline methods with respect to both $INTRA_d$ and $INTER_d$ , in both datasets of real networks and artificial networks. As explained before, a good distance metric should have a small (negative) value for $INTRA_d$ and meaningfully larger (positive) values for $INTER_d$.\\

%
%

\begin{figure}
\begin{subfigure}[b]{0.45\textwidth}
	\includegraphics{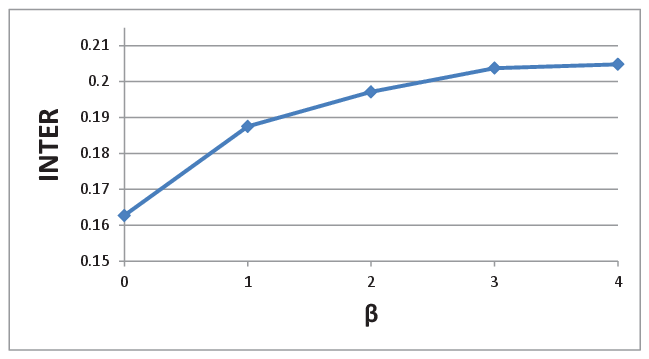}
	\caption{\label{fig:inter}}
\end{subfigure}
\begin{subfigure}[b]{0.45\textwidth}
	\includegraphics{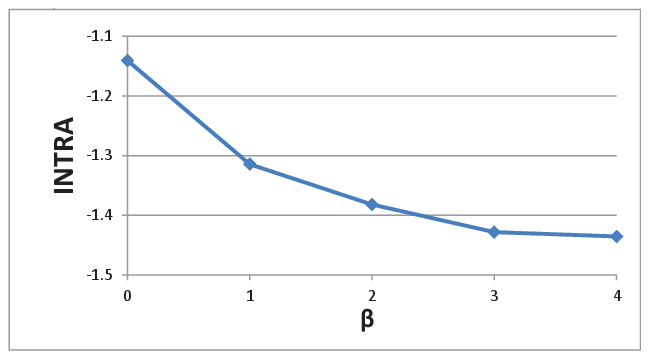}
	\caption{\label{fig:intra}}
\end{subfigure}

\caption{ \label{fig:divisionsEffect} The effect of $\beta$ parameter on the accuracy of the proposed method in artificial networks dataset, for $INTER_d$ (\subref{fig:inter}) and $INTRA_d$ (\subref{fig:intra}) criteria measures. As both figures show, larger values of $\beta$ results in better distance function with larger $INTER_d$  and smaller $INTRA_d$, but setting the $\beta$ to values larger than 3 brings no more significant improvement.}

\end{figure}

Next, we evaluate the effect of $\beta$ parameter on the accuracy of our proposed distance metric. We repeated the previous experiment for the artificial networks, but with different values of $\beta$ in the range of integer numbers from 0 to 4. As Figure \ref{fig:divisionsEffect} shows, the distance metric is improved by increasing the value of $\beta$ and it asymptotically becomes stable with values larger than $\beta =3$. This fact is consistent for both $INTER_d$  (Figure \ref{fig:inter}) and $INTRA_d$  (Figure \ref{fig:intra}) measures. According to this experiment, $\beta =3$ is an appropriate setting for this parameter as a tradeoff between the accuracy of the distance metric and the size of the quantified vector (with $\beta =3$ we will have $4\times 2^3 =32$ real numbers in the quantification of the degree distribution). So, we suggest this value for $\beta$ and we set $\beta =3$ in the rest of the experiments of this paper. It is worth reminding that even with small quantification vectors (8 numbers by setting $\beta =1$ in our previous experiments) our proposed method has outperformed the baseline distance metrics.\\

\begin{figure}
\begin{subfigure}{0.45\textwidth}

	\includegraphics{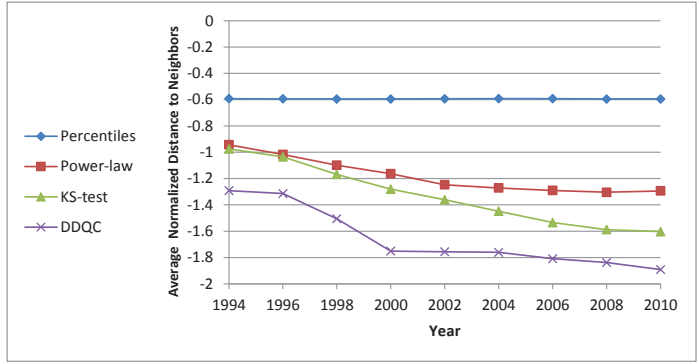}
	\caption{ \label{fig:tempcit} Normalized distance of Cit\_CiteSeerX snapshots}
\end{subfigure}

\begin{subfigure}{0.45\textwidth}
	\includegraphics{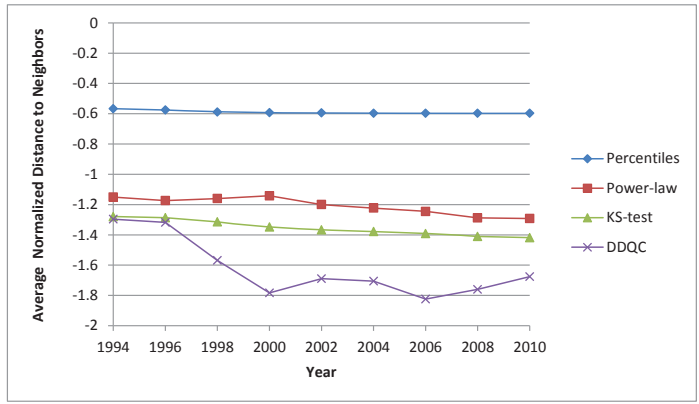}
	\caption{ \label{fig:tempcollab} Normalized distance of Collab\_CiteSeerX snapshots}
\end{subfigure}

\caption{ \label{fig:temporal} The average normalized distance of the temporal networks to their neighbor snapshots for Cit\_CiteSeerX (\subref{fig:tempcit}) and Collab\_CiteSeerX (\subref{fig:tempcollab}) temporal networks. DDQC always results in smaller normalized distances to the neighbor snapshots, which means that DDQC captures the similarity of near-in-the-time temporal networks, better than baseline methods.}

\end{figure}

In the next experiment, we consider two temporal networks extracted from CiteSeerx \cite{ref45}: Cit\_CiteSeerX citation network and Collab\_CiteSeerX collaboration network. The last version (snapshot of 2010) of these two networks already present in the real-world networks dataset. For each of the two temporal networks, we extracted nine snapshots of the network from 1994 to 2010 biannually. We then prepared two independent experiments for the two temporal networks. In each experiment, we created a union dataset by aggregating the nine instances of the temporal dataset and the real-world networks dataset. Then, we computed the pairwise distances for all the instances in the new union dataset. We normalized all the distances using Equation \ref{eq:zscore}, and then we computed the average normalized distance of each temporal network to its neighbor in the time (next and previous) snapshots. The distance of a network snapshot to its neighbor snapshots is expected to be a relatively small value. Figure \ref{fig:tempcit} and Figure \ref{fig:tempcollab} show the results of this experiment for Cit\_CiteSeerX and Collab\_CiteSeerX networks respectively. As both figures show, the average distance of a temporal network to its neighbor snapshots is smaller in DDQC. It means that DDQC captures the similarity of near-in-the-time temporal networks, better than baseline methods. \\

\begin{figure}
\begin{subfigure}[b]{0.45\textwidth}
	\includegraphics{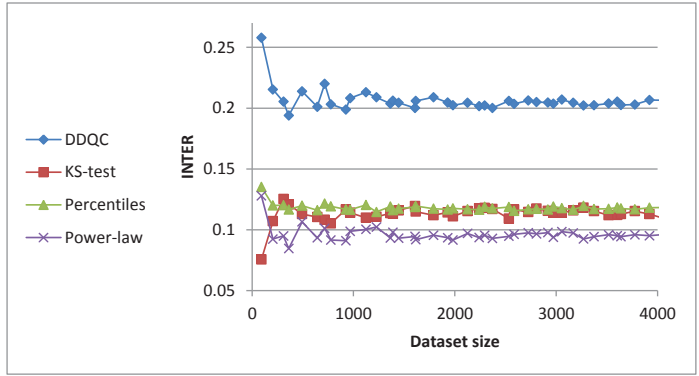}
	\caption{ \label{fig:effectofsizeinter} }
\end{subfigure}

\begin{subfigure}[b]{0.45\textwidth}
	\includegraphics{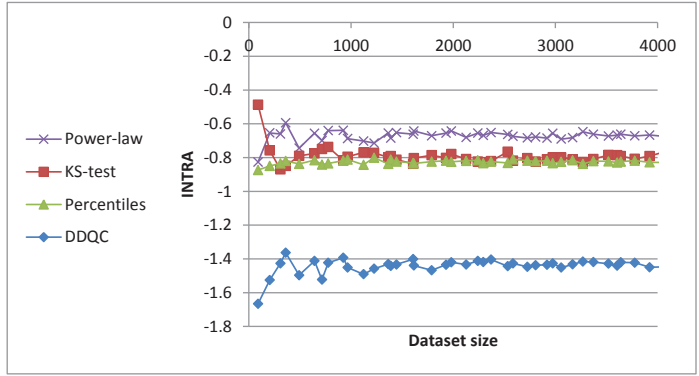}
	\caption{ \label{fig:effectofsizeintra} }
\end{subfigure}

\caption{ \label{fig:effectofsize} Effect of dataset size in stability of the evaluations. According to $INTER_d$ (\subref{fig:effectofsizeinter}) and $INTRA_d$ (\subref{fig:effectofsizeintra}) criteria measures, DDQC always outperforms the baseline methods and the results are relatively stable for dataset with more than 1,000 networks.}
\end{figure}

In our evaluations, the biggest dataset was the artificial networks with 8,000 network instances. In the next experiment, we show that the size of this dataset (8,000 network instances) is sufficient to reach a robust and stable evaluation of the distance metrics. For this purpose, we have extracted some smaller datasets from the artificial networks dataset and we calculated both $INTRA_d$ and $INTER_d$ for distance metrics over these smaller datasets. The smaller datasets have different number of networks, from 100 to 4,000 instances, and the networks are randomly selected from the original 8,000 instances of the artificial networks dataset. As Figure \ref{fig:effectofsizeinter} and Figure \ref{fig:effectofsizeintra} shows, the two measures are nearly stable for datasets with more than 1,000 networks. So, the size of the artificial networks dataset, with 8,000 network instances, is quite sufficient and the reported results are reliable. The intra/inter class distances of DDQC in Figure \ref{fig:effectofsize} show small improvements over the reported results in Figure \ref{fig:artinter} and \ref{fig:artintra}, because in this experiment we have used $\beta =3$ parameter (In the experiment reported in Figure \ref{fig:artinter} and Figure \ref{fig:artintra}, $\beta$ is equal to one).\\

\begin{figure}
\begin{subfigure}[b]{0.45\textwidth}
	\includegraphics{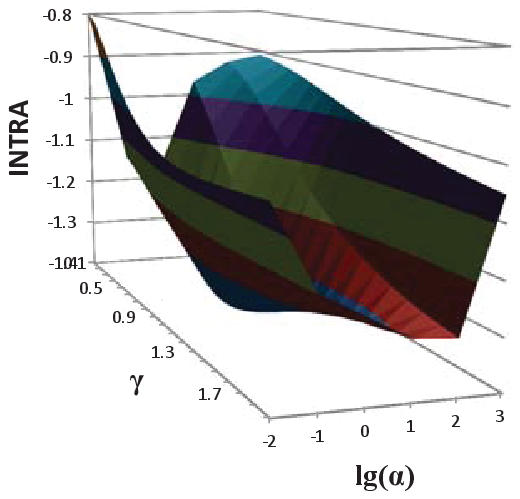}
	\caption{ \label{fig:intra3d}}
\end{subfigure}
\begin{subfigure}[b]{0.45\textwidth}
	\includegraphics{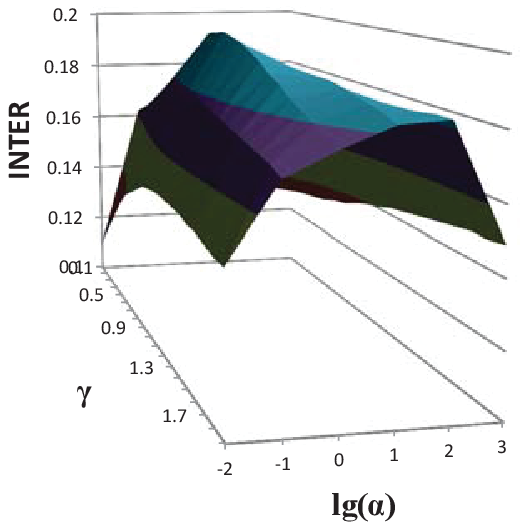}
	\caption{ \label{fig:inter3d} }
\end{subfigure}
\caption{ \label{fig:3d} Effect of $\alpha$ and $\beta$ parameters on the $INTRA_d$ (\subref{fig:intra3d}) and $INTER_d$ (\subref{fig:inter3d}) measures. The figures show a convex search space with an extremum at  $\alpha =1$ and $\gamma =0.8$ for both the diagrams.}
\end{figure}

In the last experiment, we examine different values of $\alpha$ and $\gamma$ configuration parameters to find their best settings. Five values are tested for $\alpha$ as $\alpha =\langle 2^i \rangle_{i=-2,-1,0,1,2,3}$. Setting $\alpha$ to values out of this range (i.e., $\alpha > 8$ or $\alpha < 0.25$) makes the two middle regions of the degree distribution too wide (covering almost the whole distribution) or too narrow. For $\gamma$ parameter, 20 different values are tested ($\gamma =\langle \frac{i}{10} \rangle_{i=1..20}$). Figure \ref{fig:3d} shows the average intra-class and inter-class distances of DDQC for artificial networks dataset, using the described values for $\alpha$ and $\gamma$. As Figure \ref{fig:3d} indicates, the best parameter setting is $\alpha =1$ and $\gamma =0.8$ for both the diagrams. This setting is used for the parameters in our reported experiments. The diagram indicates a convex space with no other local optimum in this search experiment. The parameters may be further tuned via a fine-grained search in the set of real numbers. Since the search space (the collection of all possible solutions) is prohibitively large, intelligent search algorithms such as genetic algorithm \cite{ref48} or simulated annealing \cite{ref49} will improve the performance of the search.\\

\section{Conclusion}
\label{Conclusion}

In this paper, we proposed a novel method for quantification and comparison of network degree distributions. The ''quantification'' is about extracting a feature vector of real numbers from a degree distribution. The ''comparison'' task is about returning a real number as the distance between two degree distributions. The distance is the counterpart of ''similarity'' and larger distances indicate less similarity.\\

To propose an appropriate distance metric for degree distributions, we first discussed the notion of distance and similarity for degree distributions according to the applications of such a comparison. The degree distribution is an indicator of the link formation process  in the network, which reflects the overall pattern of connections \cite{ref27}. Similarly evolving networks have analogous degree distributions, and we derive similarity of degree distributions according to the similarity of link formation process in the networks. For deriving the amount of similarity of networks, we introduced admissible witnesses for network similarity: similarity among the networks in three categories (same-type real networks, same-model artificial networks, and near-in-the-time temporal networks). So, we assume the networks in each of these categories have similar degree distributions. This fact is the base of our evaluations for different degree-distribution distance metrics. In our survey of existing methods for network comparison based on the degree distribution, perhaps Kolmogorov-Smirnov (KS) test is the most common method for comparing the degree distributions. But KS-test does not support quantification and needs the whole degree distribution. Power-law exponent and Percentiles \cite{ref20} are other measures for comparing degree distributions. Our proposed method, named DDQC, outperforms the existing algorithms with regard to its accuracy in various evaluation criteria. The evaluations are performed based on different criteria: The ability of the metric to classify networks, calculating inter/intra class distances, and analyzing temporal network distances over different timestamps. \\

As the future works, we will use the proposed quantification and comparison method in other application domains. Our proposed method enables the data analysis applications and data mining algorithms to employ the feature of the degree distribution as a fixed-length set of real numbers. So, it is now possible to represent a network instance with a record of features (including clustering coefficient, average path length and the quantified degree distribution) and use such records in data analysis applications. We will combine different network features along with the quantified degree distribution in an integrated distance metric for complex networks. Such an integrated distance metric will be the main building block of our future researches in evaluation and selection of network generative models and sampling methods. 

\section*{Acknowledgements}
\label{Acknowledgements}
We appreciate Sadegh Motallebi, Sina Rashidian and Javad Gharechamani for their helps in implementation and evaluation of the methods and in preparation of temporal network datasets. We also thank Masoud Asadpour, Mahdieh Soleymani Baghshah and Hossein Rahmani for their valuable comments. 

\appendix

\section{Overview of Artificial and Real-world Networks }
\label{datasetsappendix}
In this appendix, we briefly describe the network datasets of this research. The ''artificial networks'' dataset consists of 8,000 networks, which are synthesized using eight generative models. For each generative model, 1000 network instances are generated using completely different parameters (no two networks are generated with the same parameters). The number of nodes in generated networks is configured from 1,000 to 5,000 nodes, with the average of 2,936.34 nodes and 13,714.75 edges in each network instance. The selected generative models are some of the important and widely used network generation methods which cover a wide range of degree distribution structures. The generative models are described in the following, along with their configuration parameters in generation of ''artificial networks'' dataset. The values of the model parameters are selected according to the hints and recommendations of the cited original papers, along with a concern of keeping the number of network edges balanced for different models.

\begin{itemize}
\setlength{\itemindent}{-1em}
\item 
\textbf{\textit{Barab\'{a}si-Albert model (BA)}}. 
This is the classical preferential attachment model which generates scale free networks with power-law degree distributions \cite{ref7}. In this model, new nodes are incrementally added to the graph, one at a time. Each new node is randomly connected to $k$ existing nodes with a probability that is proportional to the degree of the available nodes. In the artificial networks dataset, $k$ is randomly selected as an integer number from the range $1\leqslant k \leqslant 10$.
\item 
\textbf{\textit{Copying model (CM)}}. 
This model also produces scale-free networks \cite{ref42,ref43}. At each time step of this model, a new node is added with $k$ edges. With probability $\beta$, the neighbors of the new node are selected randomly among the existing nodes, and with probability $1-\beta$, one existing node is randomly chosen and $k$ of its neighbors are ''copied'' as the ends of the new edges. In the artificial networks dataset, the $\beta$ parameter is randomly chosen from the all valid real numbers $0 < \beta <1$. 
\item 
\textbf{\textit{Erd\H{o}s-R\'{e}nyi (ER)}}. 
This model generates completely random graphs with a specified density \cite{ref44}. Network density is defined as the ratio of the existing edges to potential edges. The density of the ER networks in the artificial networks dataset is randomly selected from the range $ 0.002 \leqslant density \leqslant 0.005$.
\item 
\textbf{\textit{Forest Fire (FF)}}. 
This model, in which edge creation is similar to fire-spreading process, supports shrinking diameter and densification properties along with heavy-tailed in-degrees and community structure \cite{ref9}. This model is configured by two main parameters: Forward burning probability ($p$) and backward burning probability ($p_b$). For generating artificial networks dataset, we fixed $p_b=0.32$ and selected $p$ randomly from the range $0 \leqslant p \leqslant 0.3$.
\item 
\textbf{\textit{Kronecker graphs (KG)}}. 
This model generates realistic synthetic networks by applying a matrix operation (the kronecker product) on a small initiator matrix \cite{ref29}. The model is mathematically tractable and supports many network features including small path lengths, heavy tail degree distribution, heavy tails for eigenvalues and eigenvectors, densification, and shrinking diameters over time. The KG networks of the artificial networks dataset are generated using a $2 \times 2$ initiator matrix. The four elements of the initiator matrix are randomly selected from the ranges: $0.7 \leqslant P_{1,1} \leqslant 0.9, 0.5 \leqslant P_{1,2} \leqslant 0.7, 0.4 \leqslant P_{2,1} \leqslant 0.6, 0.2 \leqslant P_{2,2} \leqslant 0.4$.
\item 
\textbf{\textit{Random power-law (RP)}}. 
This model follows a variation of ER model and  generates synthetic networks with power law degree distribution \cite{ref30}. This model is configured by the power-law degree exponent ($\gamma$). In our parameter setting for generating artificial networks dataset, $\gamma$ is randomly selected from the range $2.5 < \gamma <3$.
\item 
\textbf{\textit{Watts-Strogatz model (WS)}}. 
The classical Watts-Strogatz small-world model synthesizes networks with small path lengths and high clustering \cite{ref5}. It starts with a regular lattice, in which each node is connected to $k$ neighbors, and then randomly rewires some edges of the network with rewiring probability $\beta$. In WS networks of the artificial networks dataset, $\beta$ is fixed as $\beta =0.5$, and $k$ is randomly selected from the integer numbers between 2 and 10 ($2 \leqslant k \leqslant 10$).
\item 
\textbf{\textit{Regular graph model (RG)}}. 
In a regular graph, each node has exactly $k$ number of neighbors. In the artificial networks dataset, $k$ is randomly selected from the range ($2 \leqslant k \leqslant 10$). 
\end{itemize}

Table \ref{tab:realnets} describes the graphs of the ''real-world networks'' dataset, along with the category, number of nodes and edges, and the source of these graphs. Most of these networks are publicly available datasets. Two temporal networks (Cit\_CiteSeerX and Collab\_CiteSeerX) are extracted from CiteSeerx digital library \cite{ref45}, using a web crawler software tool. 

\begingroup
\squeezetable
\begin{table}[htbp]
\caption{ \label{tab:realnets} Dataset of real-world networks}

    \begin{tabular}{|c|lrrr|}
    \hline
    \textbf{	Category} & \textbf{ID} & \textbf{Vertices} & \textbf{Edges} & \textbf{Source} \\
    \hline
    \multirow{4}{*}{Citation Network} & Cit-HepPh & 34,546 & 420,899 & SNAP \cite{ref50} \\
\cline{2-5}          & Cit-HepTh & 27,770 & 352,304 & SNAP \cite{ref50} \\
\cline{2-5}          & dblp\_cite & 475,886 & 2,284,694 & DBLP \cite{ref51} \\
\cline{2-5}          & Cit\_CiteSeerX & 1,106,431 & 11,791,228 & CiteSeerX [45] \\
    \hline
    \multirow{10}{*}{Collaboration Network} & CA-AstroPh & 18,772 & 198,080 & SNAP  \cite{ref50} \\
\cline{2-5}          & CA-CondMat & 23,133 & 93,465 & SNAP  \cite{ref50} \\
\cline{2-5}          & CA-HepTh & 9,877 & 25,985 & SNAP  \cite{ref50} \\
\cline{2-5}          & Collab\_CiteSeerX & 1,260,292 & 5,313,101 & CiteSeerX [45] \\
\cline{2-5}          & com-dblp & 317,080 & 1,049,866 & SNAP  \cite{ref50} \\
\cline{2-5}          & dblp\_collab & 975,044 & 3,489,572 & DBLP \cite{ref51} \\
\cline{2-5}          & dblp20080824 & 511,163 & 1,871,070 & Sommer \cite{ref52} \\
\cline{2-5}          & IMDB-USA-Commedy-09 & 4,155 & 16,679 & Rossetti \cite{ref53} \\
\cline{2-5}          & CA-GrQc & 5,242 & 14,490 & SNAP  \cite{ref50} \\
\cline{2-5}          & CA-HepPh & 12,008 & 118,505 & SNAP  \cite{ref50} \\
    \hline
    \multirow{4}{*}{Communication Network} & EmailURV & 1,133 & 5,451 & Aarenas \cite{ref54} \\
\cline{2-5}          & Email-Enron & 36,692 & 183,831 & SNAP \cite{ref50,ref55} \\
\cline{2-5}          & Email-EuAll & 265,214 & 365,025 & Konect \cite{ref56} \\
\cline{2-5}          & WikiTalk & 2,394,385 & 4,659,565 & SNAP \cite{ref50} \\
    \hline
    \multirow{7}{*}{Friendship Network} & Dolphins & 62    & 159   & NetData \cite{ref57} \\
\cline{2-5}          & facebook-links & 63,731 & 817,090 & MaxPlanck \cite{ref58} \\
\cline{2-5}          & Slashdot0811 & 77,360 & 507,833 & SNAP \cite{ref50} \\
\cline{2-5}          & Slashdot0902 & 82,168 & 543,381 & SNAP \cite{ref50} \\
\cline{2-5}          & soc-Epinions1 & 75,879 & 405,740 & SNAP \cite{ref50} \\
\cline{2-5}          & Twitter-Richmond-FF & 2,566 & 8,593 & Rossetti \cite{ref53} \\
\cline{2-5}          & youtube-d-growth & 1,138,499 & 2,990,443 & MaxPlanck \cite{ref58} \\
    \hline
    \multirow{4}{*}{Graph of Web Pages} & web-BerkStan & 685,230 & 6,649,470 & SNAP \cite{ref50} \\
\cline{2-5}          & web-Google & 875,713 & 4,322,051 & SNAP \cite{ref50} \\
\cline{2-5}          & web-NotreDame & 325,729 & 1,103,835 & SNAP \cite{ref50} \\
\cline{2-5}          & web-Stanford & 281,903 & 1,992,636 & SNAP \cite{ref50} \\
    \hline
    \multirow{4}{*}{P2P Network} & p2p-Gnutella04 & 10,876 & 39,994 & SNAP \cite{ref50} \\
\cline{2-5}          & p2p-Gnutella05 & 8,846 & 31,839 & SNAP \cite{ref50} \\
\cline{2-5}          & p2p-Gnutella06 & 8,717 & 31,525 & SNAP \cite{ref50} \\
\cline{2-5}          & p2p-Gnutella08 & 6,301 & 20,777 & SNAP \cite{ref50} \\
    \hline
    \end{tabular}%
\end{table}
\endgroup

\bibliography{shortbib}

\end{document}